\documentclass[aps, onecolumn, groupedaddress, superscriptaddress, floatfix, longbibliography, 12pt]{revtex4-2}

\usepackage{amsmath}
\usepackage{amssymb}
\usepackage{graphicx}
\usepackage{float}
\usepackage[english]{babel}
\usepackage{dsfont}
\usepackage{epstopdf}
\usepackage{soul}
\usepackage{color}
\usepackage{braket}
\usepackage[left]{lineno}

\usepackage[normalem]{ulem}
\usepackage[colorlinks=true,
            linkcolor=blue,
            urlcolor=blue,
            citecolor=blue]{hyperref}

\definecolor{SMblue}{rgb}{0.2,0.2,1}
\definecolor{SMblue2}{rgb}{0,0,0}
\definecolor{GRgreen}{rgb}{0.31,0.74,0.47}
\emergencystretch=\maxdimen
\usepackage{titlesec}
\titlespacing\subsection{0pt}{12pt plus 4pt minus 2pt}{12pt plus 2pt minus 2pt}
\titlespacing\section{0pt}{12pt plus 4pt minus 2pt}{7pt plus 2pt minus 2pt}
\titlespacing\subsection{0pt}{12pt plus 4pt minus 2pt}{7pt plus 2pt minus 2pt}

\usepackage{enumitem}

\begin{document}

\title{Observation of Unidirectional s-p Orbital Topological Edge States\\in Driven Photonic Lattices}

\author{Gayathry Rajeevan}
\affiliation{Department of Physics, Indian Institute of Science, Bangalore 560012, India}
\author{Sebabrata Mukherjee}
\email{mukherjee@iisc.ac.in}
\affiliation{Department of Physics, Indian Institute of Science, Bangalore 560012, India}
\date{\today}

\begin{abstract}
Time-periodic modulation of a static system is a powerful method for realizing robust unidirectional topological states. So far, all such realizations have been based on interactions among $s$ orbitals, without incorporating inter-orbital couplings. Here, we demonstrate higher-orbital Floquet topological insulators by introducing periodically modulated couplings between the optical $s$ and $p$ orbitals in a square lattice. The staggered phase of the $s$-$p$ couplings gives rise to a synthetic uniform $\pi$ magnetic flux per plaquette of the lattice, and periodic driving of the couplings opens a topological bandgap, characterized by the Floquet winding number. We image topological edge modes of $s$-$p$ orbitals traveling unidirectionally around a corner. Here, the topological phases are realized by a combined effect of the periodic driving and synthetic magnetic flux. Consequently, when the synthetic flux is turned off, the system becomes trivial over a range of driving parameters.  Our results open a promising pathway for exploring topological phenomena by introducing the orbital degree of freedom.
\vspace{-1.0mm}
\noindent
\end{abstract}
\maketitle

\clearpage 

\noindent {\bf Introduction}\\
Topological states were originally observed in two-dimensional electron gases under a strong transverse magnetic field, which breaks the time-reversal symmetry~\cite{klitzing1980new, thouless1982quantized}. Since then, various intriguing topological phases have been realized in a wide range of experimental settings~\cite{wang2009observation, rechtsman2013photonic, hafezi2013imaging, jotzu2014experimental, susstrunk2015observation, karzig2015topological, ningyuan2015time, wintersperger2020realization}, including ultra-cold atoms and photonics systems. Specifically, topological photonics~\cite{raghu2008analogs,lu2014topological,ozawa2019topological,smirnova2020nonlinear} has recently emerged as an important subfield of modern optics, uncovering novel physics and holding a great promise for robust  device applications~\cite{hafezi2011robust, bandres2018topological}.

{\color{SMblue2}Periodically driving a system's Hamiltonian -- known as Floquet engineering~\cite{goldman2014periodically, eckardt2017colloquium, garanovich2012light} %
-- is a powerful and convenient tool for creating topologically non-trivial materials. %
For example, photonic Chern insulators~\cite{rechtsman2013photonic} have been realized by helically modulating the waveguide paths of a  honeycomb lattice, and anomalous Floquet topological insulators, which has no static analogue, were observed in slowly-driven lattices~\cite{mukherjee2017experimental, rudner2013anomalous, maczewsky2017observation, mukherjee2023period}. 
Such Floquet topological states have been demonstrated in engineered lattice systems where each site usually supports a single state, i.e., the $s$-orbital. 
Although experimentally challenging, the incorporation of higher-orbital interactions into these driven systems %
can provide 
an additional degree of controllability in exploring novel topological phases~\cite{li2016physics, muller2007state}. 
The orbital physics plays an important role in understanding a variety of condensed matter systems, ranging from transition-metal oxides~\cite{tokura2000orbital} to orbital superfluids~\cite{wirth2011evidence}. 
In photonics, higher orbitals have been exploited in band engineering~\cite{xia2025fully, milicevic2017orbital} and to create synthetic magnetic flux~\cite{jorg2020artificial, caceres2022controlled}; see also \cite{vicencio2025multi, noh2025orbital}. %
Additionally, zero-dimensional topological edge states have been studied in static systems~\cite{li2013topological, schulz2022photonic, mazanov2024photonic} with inter-orbital couplings. 
Such couplings have also found promising applications in the design of compact photonic devices~\cite{gross2014three, birks2015photonic, rajeevan2025nonlinear}. %

In this work, we experimentally demonstrate two-dimensional higher-orbital Floquet topological insulators making use of the engineered couplings between the optical $s$ and $p_y$ (hereafter referred to as $p$) orbitals of neighboring sites. The photonic lattices are fabricated using femtosecond laser-writing~\cite{davis1996writing, szameit2010discrete}, which allows us
to fine-tune the waveguide refractive index profiles and precisely modulate the waveguide paths to enable periodically modulated $s$-$p$ orbital couplings.
We implement a four-step driving protocol where the $s$-$p$ couplings are switched on and off in a cyclic manner.
Unlike the previous works~\cite{rudner2013anomalous, mukherjee2017experimental, maczewsky2017observation} on anomalous Floquet topological insulators, 
one vertical coupling in each plaquette of the lattice is 
made negative by utilizing the odd parity of $p$ orbitals, resulting in a uniform synthetic $\pi$ magnetic flux. 
The periodic driving, along with the %
synthetic flux, realizes topological phases characterized by the Floquet winding numbers. %
Specifically, as the driving parameter is tuned, the system undergoes transitions from a Chern insulator to anomalous Floquet topological insulator and subsequently back to a Chern insulator.
In experiments, we image the topological edge modes of $s$-$p$ orbitals traveling around a corner of the lattice. 
To highlight the importance of the synthetic magnetic flux, we also demonstrate the existence of such unidirectional edge modes using certain parameters for which the system becomes trivial when the $s$-$p$ orbital couplings are replaced by $s$-$s$ couplings with vanishing flux.

\begin{figure*}[] 
    \centering
\includegraphics[width=1\linewidth]{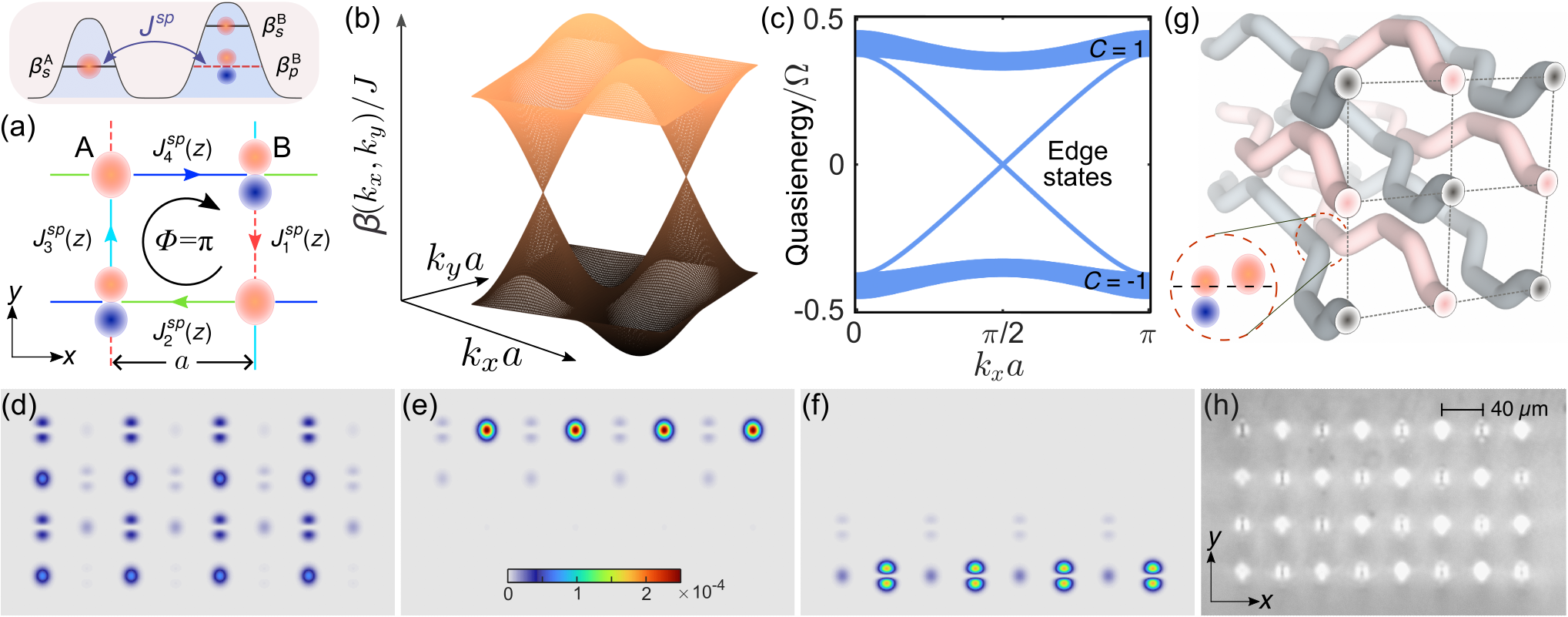}
    \caption{{Photonic Floquet topological insulator of $s$-$p$ orbitals.} (a) Schematic of a square lattice of $s$-$p$ orbitals with two sites ($\text{A}$ and $\text{B}$) per unit cell and nearest-neighbor couplings $J_{1-4}^{sp}$. Note that $J_1^{sp}$ is negative, realizing a uniform synthetic $\pi$ magnetic flux per plaquette. The inset on the top shows the refractive index profile and the resonance condition, $\beta_s^\text{A}\!=\!\beta_p^\text{B}$. 
    (b) Photonic bands  of propagation constant $\beta(k_x, k_y)$ (analogous `energy') for homogeneous coupling strengths $|J_{1-4}^{sp}|=J$, exhibiting two Dirac cones at $(k_xa,k_ya)\!=\!(\pm\pi/2, 0)$ with linear dispersion. The system in (a) is made topological by periodically switching on and off the couplings (one at a time) in a clockwise manner along the propagation distance, $z$. (c) Quasienergy spectrum for a strip geometry along the $y$ direction of the Floquet square lattice with experimentally realized parameter $\Lambda_{1-4}\!=\! 0.76\,\pi/2$. 
(d-f) Intensity distributions
of bulk and edge modes (on the top and bottom edges) of $s$-$p$ orbitals, respectively.
(g) Simplified sketch showing the photonic realization of the Floquet topological insulator of $s$-$p$ orbitals. The coupling between any two sites is switched on and off by synchronously bending the waveguide paths along $z$. Non-zero horizontal couplings
are realized by changing the waveguide positions as indicated in the inset.
(h) White-light transmission  
micrograph (cross-section) of a femtosecond laser fabricated
$s$-$p$ orbital square lattice.}
    \label{Fig_SP_model}
\end{figure*}

\vspace{0.5cm}

\noindent {\bf Results}\\
\noindent {\bf Model -- } We consider a photonic square lattice 
consisting of two sites ($\text{A}$ and $\text{B}$) 
per unit cell, %
as illustrated in Fig.~\ref{Fig_SP_model}(a). The $\text{A}$ site 
supports only the fundamental 
mode ($s$ orbital) with on-site energy $\beta_s^\text{A}$, while the
$\text{B}$ site supports both the $s$ and $p$ orbitals with on-site energies $\beta_s^\text{B}$ and $\beta_p^\text{B}$, respectively.
The $s$ orbital of $\text{A}$ sites and the $p$ orbital of $\text{B}$ sites are phase-matched (i.e., $\beta_s^\text{A}\!=\!\beta_p^\text{B}$), whereas the on-site energy of the  $s$ orbital of the B site is largely detuned. 
In this situation, an initial state localized at the $\text{A}$ site can tunnel only to the $p$ orbital of a neighboring $\text{B}$ site, while the $s$ orbital of the $\text{B}$ site remains effectively decoupled. Hence, we shall consider the tunneling of light among these $s$ and $p$ orbitals of A and B sites, respectively.

The $s$-$p$ orbital couplings 
in the lattice are modulated periodically along the propagation distance $z$ in a clockwise manner. The total driving period $z_0$ is equally divided into four steps, and during the $m$-th step, i.e., $(m-1)z_0/4\!\leq z \!\leq mz_0/4$, only $J_{m}^{sp}(z)$ couplings are switched on; $m\!=\!1,2,3$, and $4$. 
Note that each lattice site is coupled only to one of its nearest neighbor at a given $z$, and the optical power transferred from one waveguide to its nearest neighbor is determined by the parameter $\Lambda_m\!=|\!\int J_{m}^{sp}(z) \,{\text{d}}z|$, where the integral is carried over the $m$-th quarter of the driving period. %
In the scalar-paraxial approximation, the propagation of optical fields through the photonic lattice can be described by the following discrete Schr{\"o}dinger-like equation~\cite{christodoulides2003discretizing}
\begin{eqnarray}
i\frac{\partial}{\partial z} \psi_j(z)\! = \!-\beta_j \, \psi_{j} -\sum_{\left\langle j' \right\rangle} J_{m}^{sp}(z) \, \psi_{j'} \; , \label{eq1}
\end{eqnarray}
where $\psi_j$ is the envelope of the optical field at the $j$-th waveguide. %
Here, the propagation distance plays the role of time ($z\leftrightarrow t$), and the propagation constant $\beta$ of a photonic orbital is the analogous on-site energy.
Since the lobes of the $p$ orbital are $\pi$ out of phase, %
the vertical coupling $J_1^{sp}$ is negative, 
and the other three couplings $J_{2-3}^{sp}$ are positive, %
producing a uniform synthetic magnetic $\pi$ flux per plaquette of the square lattice.
In other words, when the optical field encircles a closed loop around a plaquette, it acquires a $\pi$  phase -- analogous to the Aharonov-Bohm phase~\cite{aharonov1959significance}.
In the absence of the periodic modulation, the Fourier-space static Hamiltonian $\hat{H}(k_x, k_y)$ 
of the square lattice with $\pi$ flux  %
supports two inequivalent Dirac points~\cite{dubvcek2015harper} at $(k_xa,k_ya)\! = \!(\pm\pi/2, 0)$, where $\mathbf{k} \!=\! (k_x,k_y)$ is the quasimomentum and, $a$ is the lattice constant; %
see Fig.~\ref{Fig_SP_model}(b). In this situation, the application of $z$-periodic modulation to the couplings effectively breaks {\it time} reversal symmetry, thereby opening a topological gap around the quasienergy $\varepsilon\!=\!0$, as described below. %

For our $z$-periodic photonic structure with $\hat{H}(z+z_0)\!=\!\hat{H}(z)$, the evolution operator is defined as~\cite{goldman2014periodically} 
$$\hat{U}(z)\! = \! \mathcal{T}\exp \Big[-i\int_0^{z} \text{d}z' \, \hat{H}(z') \Big],$$ where $\mathcal{T}$ is $z$-ordering.
Considering a strip-geometry aligned along the $y$ direction, and 
periodic along the $x$ direction, we calculate the quasienergy  
spectrum by diagonalizing the evolution operator $\hat{U}(z_0)$ over one complete period.  
In Fig.~\ref{Fig_SP_model}(c), the spectrum is shown for $\Lambda_{1-4}\! \equiv \!\Lambda\!=\! 0.76\,\pi/2$, where
topological edge modes appear in the quasienergy gap around $\varepsilon\!=\!0$.
In Figs.~\ref{Fig_SP_model}(d-f), we show the 
intensity distributions of bulk and edge modes (on the top and bottom edges) considering the experimentally realized system size and with periodic boundary condition in the $x$ direction. Notice the  mixing of $s$-$p$ orbitals in forming the bulk and edge states. Additionally, for our driving parameters, the $s$ $(p)$ orbitals exhibit a higher weightage in the top (bottom) edge, indicating the possibility of exciting these modes using a single-site initial state. 

The topology of our system is captured by the integer-valued Floquet winding number, which takes into account the complete $z$ evolution of the Floquet operator $\hat{U}(z)$~\cite{rudner2013anomalous}.
Specifically, the winding number for a 
quasienergy gap is equal to the number of topological edge modes in that gap. %
In the case of Fig.~\ref{Fig_SP_model}(c), the winding number for the bandgap around zero quasienergy $W_{\varepsilon = 0}$ is unity, as there exists one topological edge mode in this gap. 
The other bandgap around quasienergy $\varepsilon\!=\!\pm\Omega/2$ is trivial, as no topological edge states exist in this gap, meaning $W_{\varepsilon = \pm\Omega/2}=0$. 
The winding numbers can be related to the %
standard topological invariant, i.e., the Chern number. Indeed, a Floquet quasienergy band can also be characterized by the Chern numbers $\mathcal{C}$, which is equal to the difference of the winding numbers below and above that band.
Evidently, the Chern number  of the upper band of our driven $s$-$p$ lattice is unity.
It should be noted that, away from the high-frequency driving limit, as in our case, driven systems can support anomalous Floquet topological edge states, when the Chern numbers are zero for all of the Floquet bands~\cite{rudner2013anomalous, nathan2015topological}.

\begin{figure}[] 
    \centering
\includegraphics[width=0.75\linewidth]{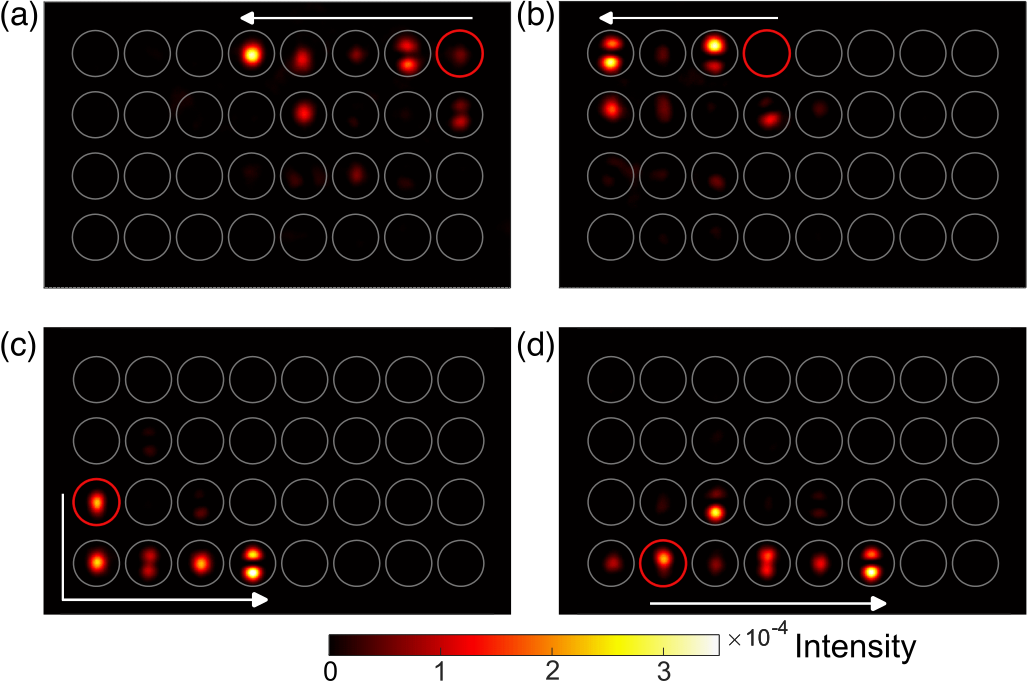}
    \caption{{Probing unidirectional $s$-$p$ orbital topological edge states.}
    Experimentally measured output intensity distributions at $z\!=\!2z_0$. Here, $\Lambda =(0.76\pm 0.02)\pi/2$.
    The red 
    circle marks the input coupling position. 
    In (a) and (b), the light is coupled to the $s$ mode of A sites at the input, whereas in (c) and (d), it is mostly ($\sim \!81\%$) coupled to the $p$ mode of B sites.
    The topological edge modes propagate unidirectionally and are not scattered by corners. Each measured intensity pattern is normalized to $1$.
    } 
    \label{Fig_SP_exp_edge_transport}
\end{figure}

\begin{figure}[] 
    \centering
\includegraphics[width=0.75\linewidth]{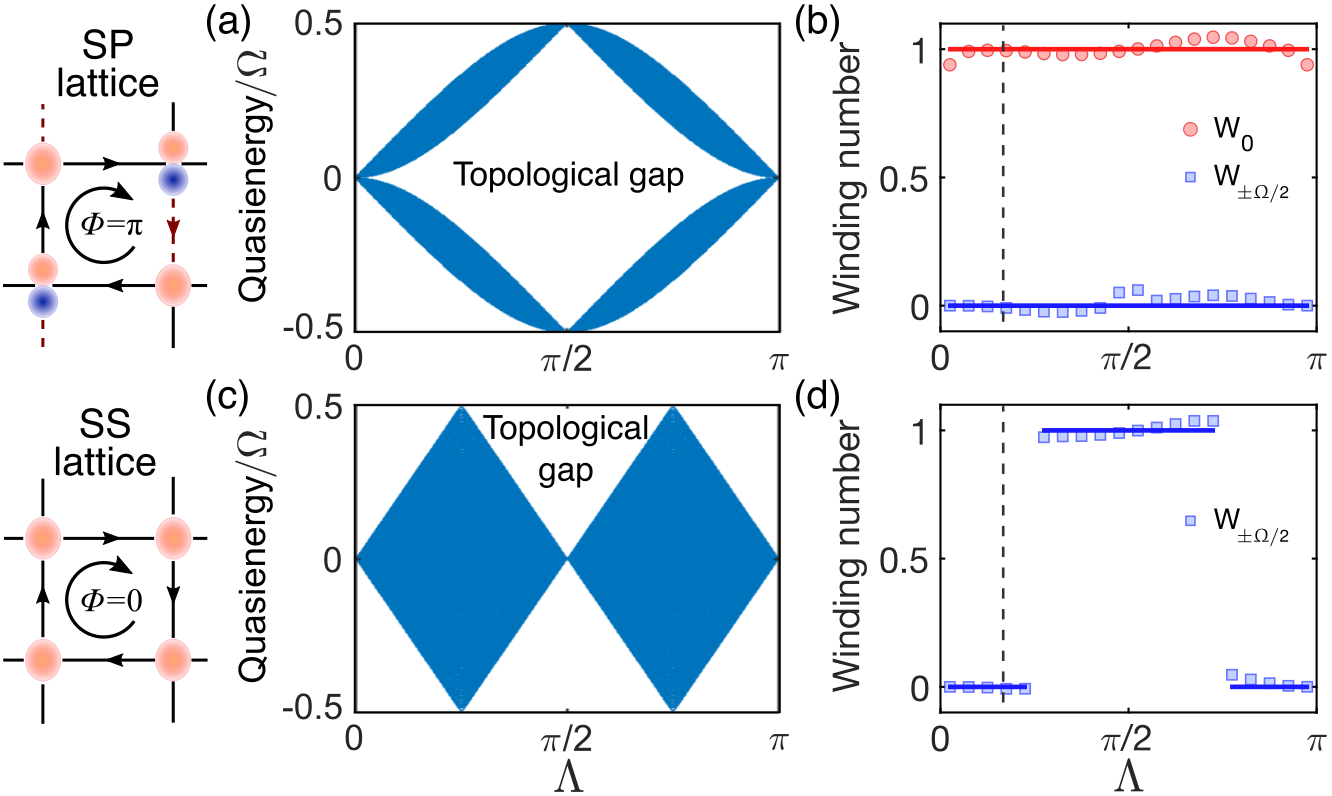}
    \caption{{Topological invariants.} (a) Bulk quasienergy spectrum %
    of the $s$-$p$ lattice as a function of $\Lambda$ showing the bandgap closing and opening, which indicates topological phase transitions. (b) Winding numbers W$_\varepsilon$ 
    for the gaps centered around quasienergy $\varepsilon\!=\!0$ and $\pm \Omega/2$ for the $s$-$p$ lattice. The solid lines are guides to the eye. (c, d) Similar calculations for a lattice with $s$-$s$ couplings with zero flux. The quasienergy gap $\varepsilon\!=\!\pm \Omega/2$ is topological for $\pi/4 < \Lambda < 3\pi/4$. The dashed vertical lines in (b, d) indicate $\Lambda$ values used in Figs.~\ref{Fig_SP_SS_edge_excitation} (a, b).
    }    \label{Fig_SP_topological_inva}
\end{figure}

 \vspace{0.5cm}

\noindent {\bf Experiments -- } 
In experiments, the coupling between any two sites is switched on and off by synchronously bending the waveguide paths as depicted in Fig.~\ref{Fig_SP_model}(g). The coupling strength decays exponentially as a function of the inter-waveguide spacing. Hence, the inter-waveguide spacing is first reduced, then kept fixed for a certain propagation distance $L$ and finally increased in a reverse manner to obtain  a step-like variation of the couplings along $z$. The $\Lambda_m$-values are controlled by varying the straight sections $L$ of the coupling. %
Since the overlap integral of the orbitals determines the inter-orbital $s$-$p$ coupling, the coupling depends heavily on the orientation of the waveguides and the supported modes~\cite{snyder1983optical}.
To achieve non-zero horizontal couplings ($J_{2, 4}^{sp}$) in the lattice, we keep the waveguides at a $45$~deg angle in the coupling region with the same orientation of the orbitals as indicated in the inset of Fig.~\ref{Fig_SP_model}(g); see also Supplementary Fig.~\ref{Fig_Supp_SP_bendycoupler}.

To experimentally probe 
the topological edge states of $s$-$p$ orbitals, modulated photonic square lattices consisting of $32$ waveguides and two driving periods
were created using fs laser-writing in a 120-mm-long BK7 glass substrate, see the white-light facet image (cross-section) in Fig.~\ref{Fig_SP_model}(h). 
The waveguides were created using a fixed laser power, and the phase-matching of orbitals was achieved by controlling the speed at which the glass material is translated through the focus of the laser beam during the fabrication; see Supplementary Sections~\ref{fab} and~\ref{phase-matching}.
All transport experiments were performed using horizontally-polarized low-power light at $980$~nm wavelength. 
Fig.~\ref{Fig_SP_exp_edge_transport} shows the topological edge states of $s$-$p$ orbitals propagating in the anti-clockwise direction. The red 
circle indicates the site where 
a Gaussian input state is launched at the input. In these experiments, single-site excitation at the A site on the top edge efficiently overlaps {(72\%)} with the topological edge states, making the output state mostly localized along the edge sites in Figs.~\ref{Fig_SP_exp_edge_transport}(a-b). 
In the case of Figs.~\ref{Fig_SP_exp_edge_transport}(c-d), %
the input state is primarily launched to the $p$ orbital of the B site by vertically shifting the position of the input beam. 
Notice that the small amount ($19\pm 4 \%$) of optical power, initially coupled to the $s$ orbital of the B site, remains localized and does not take part in the dynamics; see Supplementary Section~\ref{phase-matching}. 
The experimental observation of back-scatter-immune propagation of topological edge modes agrees well with numerical results (Eq.~\eqref{eq1}).

\begin{figure}[] 
    \centering
\includegraphics[width=0.75\linewidth]{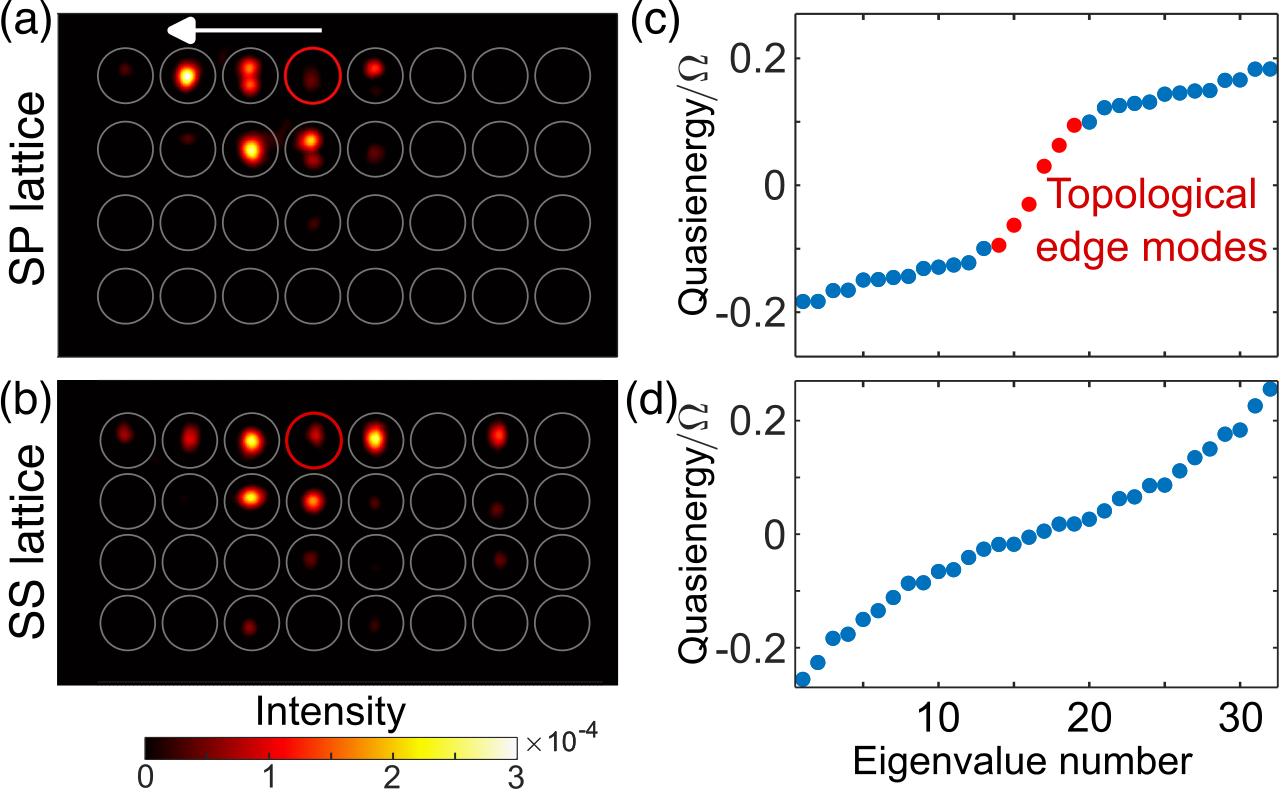}
\caption{{Importance of synthetic $\pi$ flux.} (a) Experimental signature of Floquet topological edge modes in a $s$-$p$ orbital lattice with $\Lambda\!=\!(0.3\pm0.04)\pi/2$. (b) Same as (a) but in a lattice consisting of $s$-orbitals only with $\Lambda\!=\!(0.3\pm0.03) \pi/2$. In this case, the lattice is trivial as indicated by the  near-symmetric (i.e., not unidirectional) spreading of the output intensity pattern along the edge. Each measured intensity pattern is normalized to $1$. 
(c, d) Quasienergy spectrum associated with (a, b), respectively, further highlighting the topological characteristics of the lattices}.    \label{Fig_SP_SS_edge_excitation}
\end{figure}

\vspace{0.5cm}

\noindent {\bf Importance of $\pi$ flux -- } 
Fig.~\ref{Fig_SP_topological_inva}(a) depicts the bulk quasienergy spectrum of the $s$-$p$ lattice, obtained from the $k$-space evolution operator over one complete period $\hat{U}(\mathbf{k}, z_0)$, as a function of $\Lambda$. 
Here, the bandgap closing and opening indicate topological phase transitions. For each $\Lambda$ value, the existence of chiral topological edge modes can be verified from the spectrum of a strip geometry.
Additionally, we numerically calculated~\cite{rudner2013anomalous} winding numbers considering the full $z$-evolution of $\hat{U}(\mathbf{k}, z)$ for a complete period; see Figure~\ref{Fig_SP_topological_inva}(b).
The quasienergy gap around $\varepsilon \!=\! 0$ %
hosts a topological edge mode propagating in the counter-clockwise direction for $0\!<\!\Lambda\!<\!\pi$. %
Evidently, the winding number is one (zero) for the gap around $\varepsilon \!=\!0$ $(\pm \Omega/2)$. 
Interestingly, the system becomes an anomalous Floquet topological insulator near $\Lambda\!=\pi/2$, when the quasienergy gap around $\varepsilon\!=\! \pm\Omega/2$ closes. In this case, the system is topological; 
however, the Chern number of the bulk band is zero.

To highlight the importance of the synthetic magnetic flux in realizing the above topological phases in the $s$-$p$ lattice, we now consider an $s$-$s$ lattice with vanishing flux. In other words, consider a square lattice with the same driving protocol, %
where all couplings take positive values.
In this case, the bandgap closing/opening and the topological phases are  different, as shown in Figs.~\ref{Fig_SP_topological_inva}(c, d). As $\Lambda$ is tuned, the overall bandwidth of the single bulk band changes linearly, realizing only the anomalous Floquet topological phase for $\pi/4 \!<\! \Lambda \!<\! 3\pi/4$. The winding number $W_{\varepsilon= \pm\Omega/2}$ is unity for this range of $\Lambda$.
Evidently, the presence of $\pi$ flux in the $s$-$p$ lattice not only alters the topological characteristics but also broadens the range of $\Lambda$ for which the system is topological. 
Although we focus on the $s$-$s$ lattice in Figs.~\ref{Fig_SP_topological_inva}(c, d) for experimental convenience, a similar driven square lattice with zero flux per plaquette can be realized using $s$-$p$ orbitals. This can be achieved by rotating the $p$ orbitals by $45$ deg with respect to the vertical axis.

To experimentally demonstrate this, 
we fabricated $s$-$p$ and $s$-$s$ lattices with a $\Lambda$ value near $0.3\,\pi/2$, see the vertical dashed line in Figs.~\ref{Fig_SP_topological_inva}(b, d). %
As shown in Fig.~\ref{Fig_SP_SS_edge_excitation}(a), the $s$-$p$ lattice exhibits unidirectional edge transport, along with some bulk penetration, as the single-site initial state has a relatively lower overlap ($33\%$) with the topological edge modes compared to Figs.~\ref{Fig_SP_exp_edge_transport}(a, b).
On the other hand, for the $s$-$s$ lattice, the initial state %
does not propagate unidirectionally along the edge. Instead, the optical field spreads almost symmetrically along the edge and diffracts into the bulk, indicating the absence of topological edge states, see
Fig.~\ref{Fig_SP_SS_edge_excitation}(b).
The associated quasienergy spectra for the experimentally realized parameters are presented in Figs.~\ref{Fig_SP_SS_edge_excitation}(c, d), clearly showing the topological characteristics of the two lattices. If propagated longer in larger systems, all the light in  Fig.~\ref{Fig_SP_SS_edge_excitation}(b) would travel into the bulk of the system, whereas topological edge transport along with some bulk spreading would appear in the case of Fig.~\ref{Fig_SP_SS_edge_excitation}(a). The observed intensity patterns in Fig.~\ref{Fig_SP_SS_edge_excitation} and their agreement with numerical results (see Supplementary Fig.~\ref{Fig_Supp_SP_SS_num}) clearly indicate the presence of topological edge states in the $s$-$p$ lattice with $\pi$ synthetic flux.

\vspace{0.5cm}

\noindent {\bf Discussion}\\ 
The demonstration of unidirectional $s$-$p$ orbital Floquet topological edge modes opens a new direction in the field of topological photonics. 
Specifically, accessing other higher-order orbitals and enabling novel inter-orbital coupling mechanisms~\cite{zhang2019electronically} can be of great importance. 
Higher orbitals can naturally exhibit
orbital degeneracy, strongly anisotropic couplings, and intrinsic orbital angular momentum. 
As a result, orbital degrees of freedom  
can unlock a wide range of lattice phenomena inaccessible to $s$-orbital–only models.
Additionally, the nonlinear interactions~\cite{rajeevan2025nonlinear, krupa2017spatial, wu2019thermodynamic, zhang2023realization, mukherjee2021observation} are largely influenced by the spatial extent of the photonic orbitals, making such systems well-suited for exploring  topological phenomena in the presence of engineered nonlinearity.
\\

{\it Acknowledgments.$-$} We thank Bhoomija Chaurasia,  Diptiman Sen,  and Abhinav Sinha for helpful discussions. S.M.~gratefully acknowledges support from the Ministry of Education, Government of India, through the STARS program (MoE-STARS/STARS-2/2023-0716); the Indian Institute of Science (IISc) through a start-up grant; and the Infosys Foundation, Bangalore. G.R.~thanks IISc for a Ph.D.~scholarship. 
\\ \vspace{-0.5cm}

{\it Data availability.$-$} The data supporting the findings of this study are included within the article and its supplementary information. Raw datasets are available from the corresponding author upon reasonable request.\\ \vspace{-0.5cm}

{\it Competing financial interests.$-$} The authors declare no competing financial interests.

\bibliography{SP_ref.bib}
\clearpage

\onecolumngrid
\appendix


\newcommand{\beginsupplement}{%
        \setcounter{equation}{0}
        \renewcommand{\theequation}{S\arabic{equation}}%
        \setcounter{figure}{0}
        \renewcommand{\thefigure}{S\arabic{figure}}%
         }
 
 \beginsupplement


\section*{\large {Supplementary Information} \\
\vspace*{-0.3cm}}




\section{Fabrication of photonic $s$-$p$ lattices}\label{fab}
The photonic $s$-$p$ lattices consist of optical waveguides with three-dimensionally modulated paths to implement the desired $z$-dependent couplings described in the main text, see Fig.~\ref{Fig_SP_model}. We employ femtosecond laser writing to fabricate the lattices inside a $120$-mm-long borosilicate (BK7) glass material, Fig.~\ref{Fig_Supp_SP_bendycoupler}(a).
To this end, we use circularly polarized $260$~fs (FWHM) laser pulses at $1030$~nm central wavelength, $500$~kHz repetition rate, and {$380$~nJ} pulse energy. The laser beam was focused inside the glass material mounted on high-precision $x$-$y$-$z$ translation stages (Aerotech).
As mentioned in the main text, each single-mode waveguide (A site) in the lattice was inscribed by translating the glass material twice through the focal region of the laser beam at a speed of $4$~mm/s. %
For the two-mode waveguides (B site), we increase the refractive index profile. The same two-scan method was employed to fabricate the B sites, with a vertical scan-to-scan separation of $5\,\mu$m. To precisely tune the refractive index of the B site, we vary the translation speeds.
The optimal translation speeds for the lower and upper scans were found to be $1$~mm/s and $2$~mm/s, respectively, to satisfy the phase-matching condition, $\beta_s^\text{A}\!=\!\beta_p^\text{B}$.

At the input ($z\!=\!0$) of the lattice, the waveguides are well separated ($d_x \!=\!40 \, \mu$m and $d_y\!=\!45 \,\mu$m) such that the inter-waveguide couplings are negligible. 
Specifically, the neighboring waveguides are separated more along $y$ (vertical) compared to the $x$ direction,
since the supported orbitals are elongated along the vertical direction. To switch on the $J_{1,3}^{sp}$ couplings, we first reduce the inter-site vertical spacing between the corresponding waveguides by synchronously bending them; see Figs.~\ref{Fig_Supp_SP_bendycoupler}(b, c). Then the inter-site spacing is kept fixed for
a certain propagation distance $L$ and finally increased in a reverse manner to obtain a step-like variation of the couplings along $z$. As indicated in Fig.~\ref{Fig_SP_model}(b), 
due to the $\pi$ phase difference between the lobes of the $p$ orbital, the $J_{1}^{sp}$ coupling is negative.
The $J_{2,4}^{sp}$ couplings are varied in a similar manner; however, the waveguides are kept at $45$-deg angle in the coupling region with the same
orientation of the orbitals; see Fig.~\ref{Fig_Supp_SP_bendycoupler}(d). This specific layout of the waveguides is important to implement positive-valued horizontal couplings.

\begin{figure}[t] 
    \centering
\includegraphics[width=0.85\linewidth]{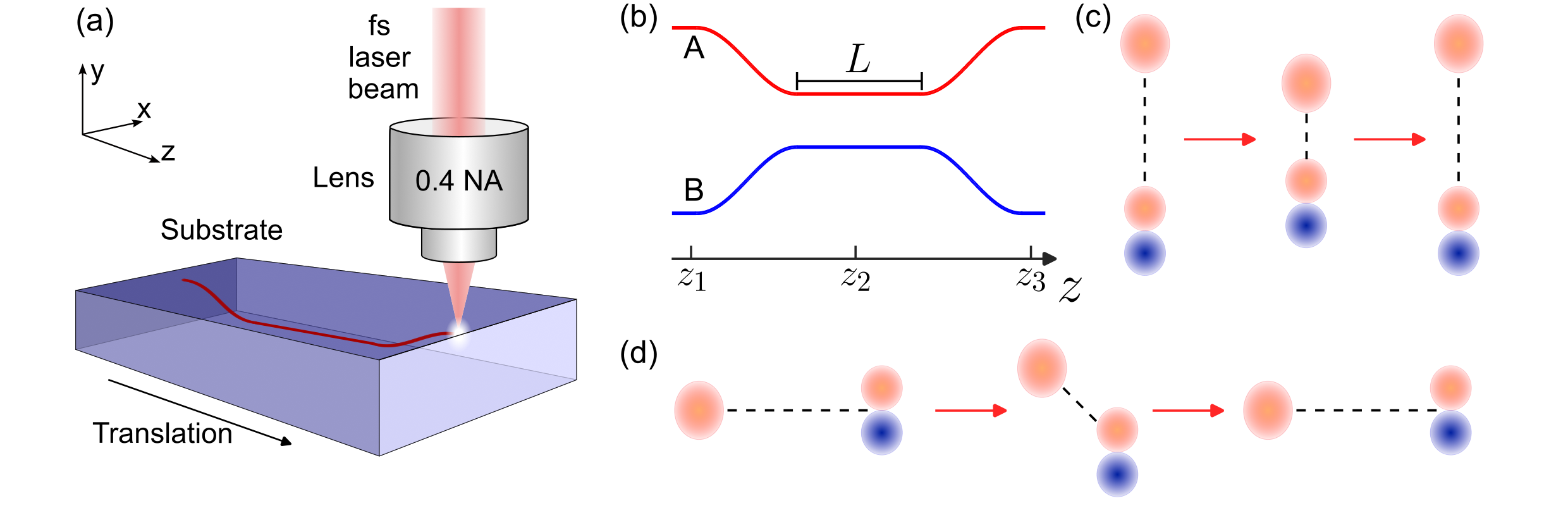}
    \caption{{(a) Schematic of femtosecond laser writing. (b) Sketch showing the bending profiles of the A and B waveguides along the propagation direction.} (c, d) Schematics of the 
    cross-sections of relative positions of the $s$ and $p$ orbitals at three different $z$ values (i.e., $z_1$, $z_2$, and $z_3$)
    for the $J_3^{sp}$ and $J_4^{sp}$ couplings, respectively.}
    \label{Fig_Supp_SP_bendycoupler}
    \end{figure}

\begin{figure}[t] 
    \centering
\includegraphics[width=0.75\linewidth]{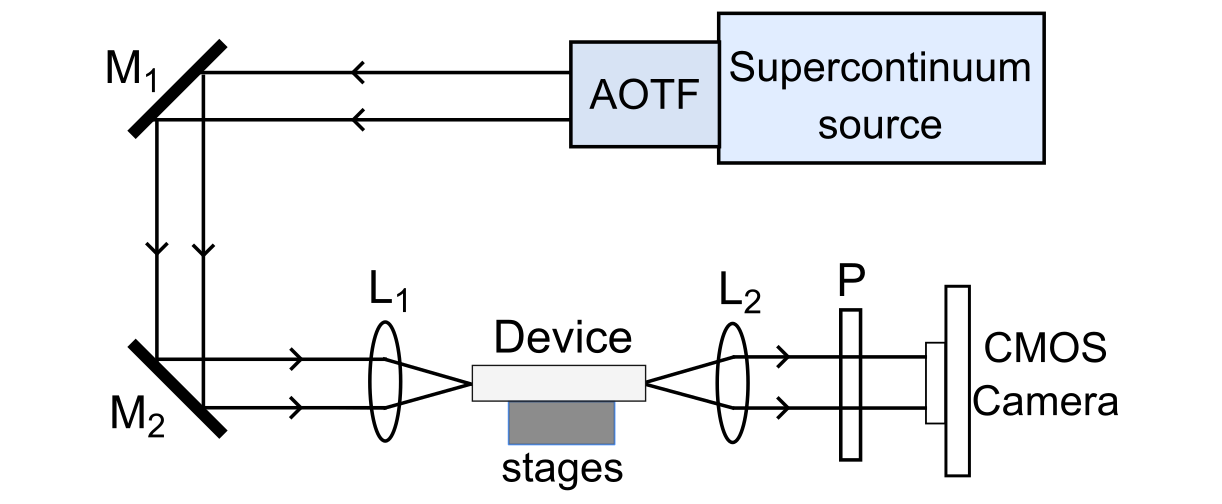}
    \caption{Schematic of the experimental characterization setup. A supercontinuum source, along with an acousto-optic tunable filter (AOTF), is used to generate wavelength-tunable collimated beam. Here, M$_{1,2}$ are silver-coated mirrors, L$_{1,2}$ are convex lenses, and P is a polarizer. For precise alignment, the photonic device is mounted on $4$-axis stages with angular and translational control, and the two lenses are mounted on $3$-axis translation stages.}
    \label{Fig_Supp_expsetup}
    \end{figure}
   
\section{Achieving the phase-matching condition}\label{phase-matching}
In the main text, we mentioned that the $s$ orbital of A sites and the $p$ orbital of B sites have the same propagation constants, i.e., $\beta_s^\text{A}\!=\!\beta_p^\text{B}$. Here, we provide experimental evidence to support this claim. %
The inset at the top of Fig.~\ref{Fig_SP_model}(a) 
shows the schematic of the refractive index profiles (scaled by the free-space wavevector $k_0$) %
of A and B sites. 
It is possible to control the waveguide refractive index contrast by either tuning the  average laser power or the translation speeds of fabrication. For our purposes, we optimize the translation speeds while keeping the fabrication power fixed.
To estimate the difference of $\beta_s^\text{A}$ and $\beta_p^\text{B}$, we fabricated seven sets of bendy $s$-$p$ couplers 
with a variable interaction length $L$. 
These devices were characterized using horizontally-polarized 
low-power light at $980$~nm wavelength, generated by a supercontinuum source; see Fig.~\ref{Fig_Supp_expsetup}. 
The near-Gaussian beam emitting from the source is focused %
to the $s$ orbital at the input, and the output powers at the $s$ and $p$ orbitals ($P_s$ and $P_p$) are measured as a function of $L$. As shown in Fig.~\ref{Fig_Supp_SP_beta}(a), the optical power from the $s$ orbital completely transfers to the $p$ orbital of the neighboring site, confirming the phase-matching condition $\beta_s^\text{A}\!=\!\beta_p^\text{B}$. The phase-matching condition in our devices were found to be wavelength dependent for a given set of fabrication parameters.

\begin{figure}[t] 
    \centering
\includegraphics[width=0.75\linewidth]{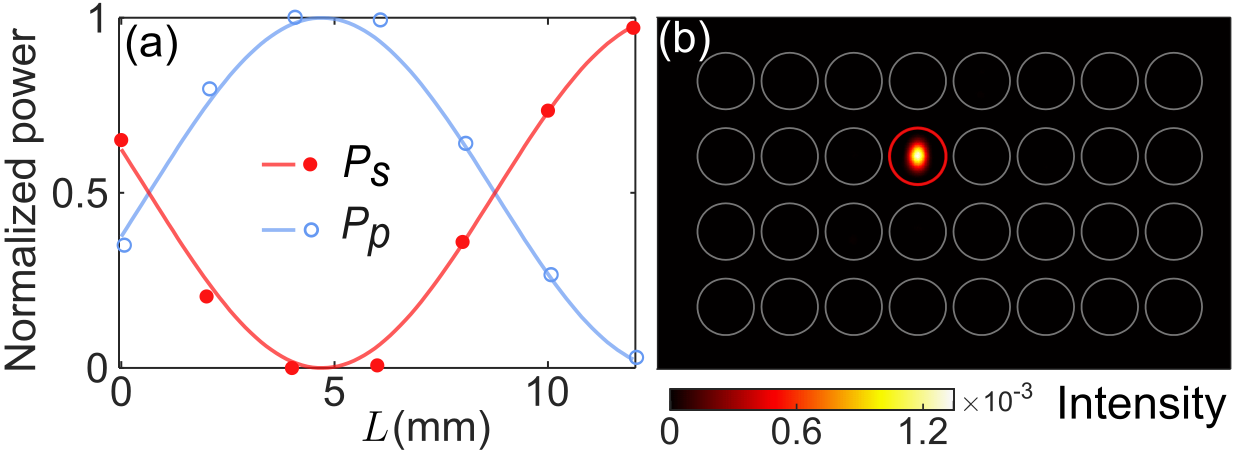}
    \caption{(a) Normalized output power in the $s$ and $p$ orbitals
($P_s$ and $P_p$) as a function of propagation distance $L$. The solid lines are obtained numerically. The complete transfer of light from the $s$ to the neighboring $p$ orbital (near $L\!=\!4.75$~mm) confirms the 
phase-matching condition, $\beta_s^\text{A}\!=\!\beta_p^\text{B}$.
(b) Experimentally measured output intensity distribution in the $s$-$p$  lattice when light is launched at the $s$ orbital of the B site (red circle). The absence of tunneling into the lattice confirms that the propagation constant detuning of this mode dominates over the couplings, i.e., $(\beta_s^{\text{B}}-\beta_p^{\text{B}})/J_{sp}\! \gg \!1$. }
    \label{Fig_Supp_SP_beta}
\end{figure}

\begin{figure}[t] 
    \centering
\includegraphics[width=0.75\linewidth]{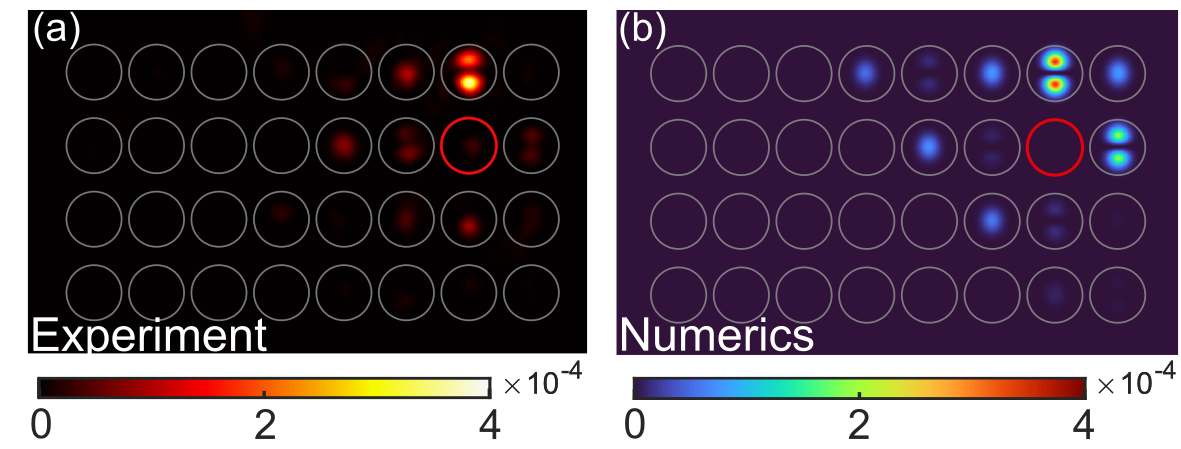}
    \caption{(a) Experimentally measured output intensity distribution in the $s$-$p$  lattice when light is coupled into the $s$ orbital of the A site (red circle). (b) Numerically obtained intensity pattern corresponding to (a).} 
    \label{Fig_Supp_SP_lattice_lambad_est}
\end{figure}

\begin{figure}[] 
    \centering
\includegraphics[width=0.75\linewidth]{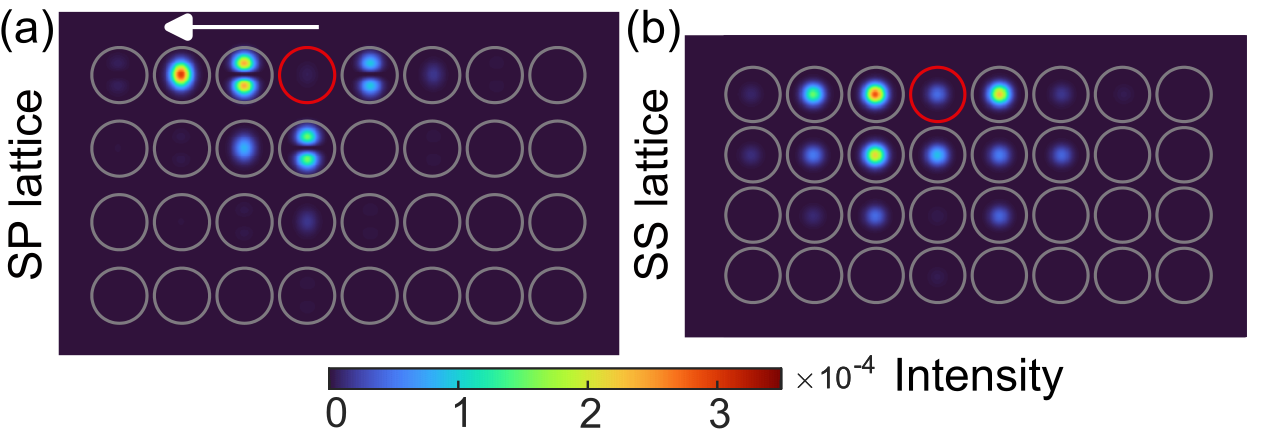}
    \caption{Numerically computed output intensity distributions in the lattice associated with Fig.~\ref{Fig_SP_SS_edge_excitation}, 
    providing an estimated value of $\Lambda\!=\!0.3\,\pi/2$.} 
    \label{Fig_Supp_SP_SS_num}
\end{figure}

As mentioned before, the $s$ orbital of B site is largely detuned, i.e., $\beta_s^\text{B} - \beta_s^\text{A}\! \gg \! J^{sp}$. To validate this experimentally, we launch light at the $s$ orbital of the B site and observe that the light does not spread in the lattice; see Fig.~\ref{Fig_Supp_SP_beta}(b). This observation also implies that the tunneling of light from the $s$ orbital of the A site to the $s$ orbital of the B site would be negligible. In other words, the dynamics of optical fields in our photonic lattices can be effectively described by the discrete Schr{\"o}dinger-like equation Eq.~\eqref{eq1} containing the $s$-$p$ orbital couplings.

\section{Estimation of $\Lambda$} \label{Lambda}
As highlighted in Fig.~\ref{Fig_SP_topological_inva}, the topology of our $s$-$p$ orbital photonic lattices is captured by the parameter 
$\Lambda_m\!=|\!\int J_{m}^{sp}(z) \,{\text{d}}z|$, where the integral is carried over the $m$-th quarter of the driving period.
We estimate the $\Lambda$-values by comparing the experimental and numerical  output intensity distributions in the lattices. For example, Fig.~\ref{Fig_Supp_SP_lattice_lambad_est}(a) presents the intensity distributions for input excitation at a bulk site. The associated numerical result obtained by R-squared maximization is shown in Fig.~\ref{Fig_Supp_SP_lattice_lambad_est}(b). By considering four such data sets with different initial excitations, the mean and standard deviation of $\Lambda$ were found to be $0.76\, \pi/2$ and $0.02 \, \pi/2$, respectively.
The numerical results associated with the second set of $s$-$p$ and $s$-$s$ lattices in Figs.~\ref{Fig_SP_SS_edge_excitation}(a, b) are presented in Fig.~\ref{Fig_Supp_SP_SS_num}. In this case, the $\Lambda$-values were found to be $0.3\,\pi/2$ with a standard deviation of $0.04 \,\pi/2$.

\end{document}